\documentstyle[aps,preprint]{revtex}
\begin{document}
\draft
\preprint{}
\title
{
Scaling near random criticality
in two-dimensional Dirac fermions
}
\author{Y. Morita and Y. Hatsugai}
\date {May 27}
\address
{
Department of Applied Physics, University of Tokyo,
7-3-1 Hongo Bunkyo-ku, Tokyo 113, Japan
}
\maketitle
\begin{abstract}
Recently
the existence of a random critical line
in two dimensional Dirac fermions is confirmed.
In this paper, we focus on
its scaling properties,
especially in the critical region.
We treat Dirac fermions in two dimensions
with two types of randomness,
a random site (RS) model and
a random hopping (RH) model.
The RS model belongs
to the usual orthogonal class and
all states are localized.
For the RH model, 
there is an additional symmetry expressed by
${\{}{\cal H},{\gamma}{\}}=0$.
Therefore, although all non-zero energy states localize,
the localization length diverges at the zero energy.
In the weak localization region,
the generalized Ohm's law
in fractional dimensions, $d^{*}(<2)$,
has been observed for the RH model.
\end{abstract}
\narrowtext
The study of
quantum phase transitions
driven by randomness has a long history.
In 1958, Anderson discussed absence of diffusion
in random systems\cite{anderson}.
In 1979, scaling arguments by Thouless et al. \cite{th}
were further developed\cite{aalr}.
The dimensionless conductance $g(L)$
is treated as the only scaling parameter,
where $L$ is the linear system size.
When
the randomness is weak ($g\gg 1$),
the metallic behavior of the $d$-dimensional system implies
$g(L)={\sigma}{L^{d-2}}$ with conductivity $\sigma$
(``the Ohm's law'').
On the other hand, when the randomness is strong ($g\ll 1$),
the wavefunction is exponentially localized,
which brings $g=g_{0}\exp(-L/{\xi})$.
In the scaling argument,
the beta function
$\beta=d\log g/d\log L=\beta (g)$
plays a central role.
In the above two asymptotic limits,
the explicit forms
are given by
$\beta(g)=(d-2)+c_{1}/g+O(1/g^{2})\ (g\gg 1)$
and
$\beta(g)=\log (g/{g_{0}})\ (g\ll 1)$.
It means that all states are localized and
the quantum phase transition is absent
in one and two dimensions.
However
random critical points can appear in two dimensions 
and much effort has been devoted to the study.
One of the cases is
quantum Hall systems where the time-reversal symmetry is broken
and they belong to the unitary class
\cite{huck}.
In spite of the
experimental and theoretical implications,
since it is beyond the weak-coupling regime,
it is still difficult to understand the critical phenomena.
In this paper, 
another example of the random criticality in two dimensions
is studied in detail
which has been discovered recently.
The properties on the critical point ( $E=0$ )
has been studied intensively \cite{recent}.
Here the scaling properties and 
the weak localization effect
are studied 
in a critical region ( $|E|{\gtrsim}0$ ). 
This is a quantum critical phenomena near the random critical line.
It brings novel weak localization effect
('Ohm's law in fractional dimensions').

Recently,
possible existence of a random critical line
in two dimensions was suggested \cite{ludwig} and
Dirac fermions with link-type randomness in two dimensions
were studied numerically \cite{lan,house}.
The random Dirac fermions were realized on a square lattice
by the $\pi $-flux model \cite{pi} with random hopping.
Our model preserves the time-reversal symmetry and
belongs to the orthogonal class.
The zero-energy states
do not localize but become critical,
which can be a prototype of critical states in two dimensions.
The density of states is $\sim |E|$ without randomness,
and becomes $\sim |E|^{\alpha}$ in the presence of randomness.
The singularity is closely  related to
the appearance of critical states.
Similar phenomenon were found in different models \cite{gw,hsw},
where the density of states has a singularity and
non-localized states appear in two dimensions.
The exponent $\alpha$ changes with strength of the randomness.
It implies the existence of the random critical line,
which is comparable with
other critical lines as
the Tomonaga-Luttinger liquid
in one-dimensional quantum systems.
The stability of the zero-energy states
against the random hopping
is due to an additional symmetry of the Hamiltonian.
The random hopping preserves the symmetry
in contrast to the site-type randomness.
This symmetry appears as
a sign change of the Hamiltonian under
the transformation $c_{j}\rightarrow (-1)^{j_{x}+j_{y}}c_{j}$.
The corresponding symmetry in an effective field theory is
denoted by ${\{}{\cal H},{\gamma}{\}}=0$
with a $4\times 4$ matrix $\gamma $\cite{lan,house}.
One of the possible scenarios proposed in \cite{house}
is that all non-zero energy states are localized and
the zero-energy states are just on the critical point.
Unfortunately,
since the localization length near the zero energy
is very large and
beyond numerically available system sizes,
no direct argument has never been given for the scenario.
In this paper,
we treat
Dirac fermions with two types of randomness
in the light of the scaling.
Support for the above scenario
is given and
weak localization effect near the random criticality
is discussed.

We study Dirac fermions
with two types of randomness
(i) random site (RS) model
and
(ii) random hopping (RH) model.
The Hamiltonian is given by
\begin{equation}
H_{\it pure}=
\sum _{i=j{\pm}(1,0)}(-1)^{j_{y}}c_{i}^{\dagger}t_{ij}c_{j}
+
\sum _{i=j{\pm}(0,1)}c_{i}^{\dagger}t_{ij}c_{j}
+
\sum _{i}c_{i}^{\dagger}V_{i}c_{i}
\label{ham}
\end{equation}
with
(i) RS model:
$t_{ij}=1$,
$V_{i}=R(W_{1})$
and
(ii) RH model:
$t_{ij}=t_{ji}=1+R(W_{2})$,
$V_{i}=0$,
where
$R(W)$'s are uniform random numbers
between $[-W,W]$.
Although we present data with $W_{1}=2.3$ and $W_{2}=1.0$,
the qualitative feature does not depend on
the strength of randomness apart from finite-size effects.

In the absence of randomness
i.e.
$t_{ij}=1$
and $V_{i}=0$,
the model is a tight-binding model
with half a flux ('$\pi$ flux') per plaquette \cite{pi}.
There are two energy bands
on the magnetic Brillouin zone
$[-{\pi},{\pi})\times[0,\pi)$,
which touch at two momenta.
Near the two momenta ${\bf k}^{i}\ (i=1,2)$,
where the energy gap closes,
the low-lying excitations are described
by massless Dirac fermions in two dimensions.
The effective Hamiltonian is given by
$
{\cal H}_{pure}=
2i\int d{\bf x}\
\Psi ^{\dagger}({\bf x})
{[}
(
\sigma _{3}\otimes \sigma _{1}
)
{\partial }_{x}
+
(
I\otimes \sigma _{3}
)
{\partial }_{y}
{]}
\Psi ({\bf x}),
$
where $\Psi ({\bf x})$ is a four component spinor.
When the Fermi energy lies at zero energy, that is,
all the negative-energy eigenstates are filled,
the Hall conductivity $\sigma _{xy}$ is ill defined.
The sign of mass determines the $\sigma _{xy}$
in the continuum theory\cite {diracsigma1,diracsigma2,diracsigma3}.
There is also the following subtlety in the tight-binding model
with half a flux ('$\pi$ flux') per plaquette
and the next-nearest-neighbor hopping $t'$ \cite{nnn}.
The $\sigma _{xy}$ is given by $t'/|t'|$ and,
when $t'=0$, the system is on the transition point
between states with different quantum  Hall conductivity.
It implies that
the zero mode carries non-zero Hall conductivity.

Here we briefly review some properties
of the above two random systems.
The effective field theory of the RS model was discussed
by mapping to the nonlinear $\sigma$ model\cite{site1}.
It predicts the localization of all states
and finite density of states at zero energy.
Recently,
whether the density of states at the zero energy
is finite or not
for random Dirac fermions,
is controversial
\cite{dos1,dos2}.
A similar model to the RS model
with dilute and strong impurities (unitary limit)
was also discussed and
consistent results with the effective field theory
were obtained\cite{d-wave,site2}.
All those results suggest that
the RS model belongs to
the usual orthogonal class
and
standard scaling arguments of
the Anderson localization \cite{aalr}
seem to be valid for the RS model.
On the other hand, in the case of the RH model,
non-localized states were discovered
at zero energy\cite{lan} and
the density of states
vanishes at the zero energy as $\sim |E|^{\alpha}$ \cite{house}.
It is not only a critical point
but also forms a random critical line,
since the exponent $\alpha$ changes with the strength of randomness.
In  Ref. \cite{ludwig},
the appearance of negative $\alpha$ was suggested
for sufficiently strong randomness.
However, even for strong randomness $W/t=1.0$, the exponent
is still positive and
the negative $\alpha$ was not observed \cite{house}.
Our model may be a part of
the 'longer' critical line.
This is analogous to
the massless phase of spinless fermions with nearest-neighbor interactions,
which is a part of the critical line
called the Tomonaga-Luttinger liquid.
It may be possible to construct models with negative $\alpha$
based on our model \cite{future}.
Divergence of the localization length near the zero energy
was also suggested numerically.
It is not allowed in a standard scenario of
the two-dimensional Anderson localization.

Let us first discuss the density of states
$\rho(E)=\langle 1/L^{2}\sum_{i}\delta(E-E_{i})\rangle$.
We diagonalize the Hamiltonian of the RS model
for finite squares of size
$L^{2}=30^{2}$, $40^{2}$ and $50^{2}$
and
ensemble average over $10000$, $5000$ and $3360$ realizations
is performed respectively.
The finite size effect is small
for the density of states and
only the result for $L^{2}=50^{2}$ is shown in Fig. 1.
The result for the RH model is also shown for comparison\cite{house}.
Finite density of states at zero energy
is created for the RS model, ${\rho}(E=0)\neq 0$.
It is in contrast to the RH model,
where the density of states
vanishes as $\sim |E|^{\alpha}$, ${\rho}(E=0) = 0$.
The difference may be related to the presence of random criticality 
in the RH model, which will be discussed later.

Next
let us discuss
scaling properties of the Thouless number
$g(E)={V(E)}/{\Delta(E)}$,
where $V(E)$ is an energy shift
obtained by replacing
periodic boundary condition with
antiperiodic boundary condition
in one direction
and $\Delta(E)$ is a local mean level spacing
near the energy $E$.
The Thouless number $g(E)$
tells us how the wavefunction is extended in the space.

Numerical results for the $g(E)$
are shown in Fig. 2.
They
are shown for $L^{2}=50^{2}$ and
average within an energy window
is also performed together with the ensemble average.
The results suggest that, in both cases, 
the localization length grows near the zero energy.
The difference is that the growth is 'singular' for the RH model,
which is related to the presence of random criticality in the RH model.
Although the difference is clear
between the RS model and the RH model,
it is crucial to apply scaling arguments
to obtain definite results,
which will be given below.

In the following,
we assume $g=g(L,E,W)=F(y=L/\xi(E,W))$
with the localization length $\xi$,
which means $g(L,E,W)$'s with different $L$ and $E$
are on a single smooth curve $g=F(y)$ 
using the localization length ${\xi}(E,W)$ 
(scaling hypothesis)\cite{aalr}\cite{2sc}.
We assume a functional form of the $\xi$ as
\begin{equation}
\xi (E,W)
=
|E|^{\beta(W)}{\bar \xi}(|E|,W)
\end{equation}
with a smooth function
${\bar \xi}({\epsilon},W)
=1+\xi _{1}(W){\epsilon}+\cdots$.
The $\xi _{n}$'s are chosen so that
$g(L,E,W)$'s with different $L$ and $E$
are on a single smooth curve $g=F(y)$.
Here
the localization length $\xi(E,W)$ is
introduced to define a dimensionless parameter
$y(=L/{\xi}(E,W))$
and is determined by the scaling hypothesis.
It is related to the usual localization length $\xi_{\rm loc}(E,W)$ of
the exponentially localized wavefunction
$|\psi ({\bf x},E,W)|\sim \exp (-|{\bf x}-{\bf x}_{0}|/\xi_{\rm loc}(E,W))$
as $\xi_{\rm loc}(E,W)=c(W)\xi(E,W)$.
Fitting our numerical results, we obtain
\begin{eqnarray}
\beta &=&0 
{\rm \ for\ the\ \ RS\ model\ (Fig.\ 3)},
\nonumber
\\
\beta &=&-0.75 
{\rm \ for\ the\ \ RH\ model\ (Fig.\ 4)},
\nonumber
\end{eqnarray}
where $W_{1}=2.3$ for the RS model 
and $W_{2}=1.0$ for the RH model.
This implies that
(i) RS model:
all states are localized,
and
(ii) RH model:
all non-zero energy states are localized
with the localization length $\xi (E,W)$
which is diverging as $E\rightarrow 0$, and
the zero-energy states are just on a critical point.

Let us discuss the above results
in the light of the scaling.
Assume that
the beta function for the Thouless number $g$
obeys the scaling form i.e.
$\beta=d\log g/d\log L=\beta (g)$.
For an almost metallic state i.e. $g\gg 1$,
we expect that $\Delta(E)\sim 1/L^{d}$
and
$V(E)\sim 1/L^{2}$ due to the level repulsion
(note that
$V(E)\sim 1/L$ for pure systems,
since there is no level repulsion).
Therefore
$g\sim L^{d-2}$ and
$\beta(g)=(d-2)+c_{1}/g+O(1/g^{2})$ for large $g$.
In particular, for $d=2$,
$\beta(g)=c_{1}/g+O(1/g^{2})$ and
$g\sim \ln (L/{\xi})$ for large $g$.
We confirmed that
the results of the RS model can be fitted to this form,
which is consistent
with usual scaling arguments
of the Anderson localization
\cite{aalr}(see Fig. 3).
On the other hand,
although
states near zero energy in the RH model
are localized in the thermodynamic limit,
they behave as critical states
due to the large localization length
beyond the available system size.
Then
we can expect
the $g$ behaves as
$\sim (L/{\xi})^{\gamma}\ ({\gamma}<0)$
for the available system size.
The results for the RH model
are consistent with this discussion (see Fig. 4).
Note that,
since $g\sim \exp (-L/{\xi})$ for large $L/{\xi}$,
there is a correction in $L/{\xi}$ to the above expression
which is assumed to be
$g=(L/{\xi})^{\gamma}(g_{0}+g_{1}L/{\xi}+\cdots)$.
We obtained good agreement
with this consideration for the RH model.

Here we define an anomalous dimension $d^{*}$ as
\begin{equation}
d^{*}=\lim _{g\rightarrow \infty}\beta (g)+2={\gamma}+2<2
\end{equation}
for states in a critical region in two dimensions.
It means $g{\simeq}\sigma L^{d^{*}-2}$
in the weak localization regime ($g\gg 1$).
This is a generalized Ohm's law in fractional dimensions
between 1 and 2.
For example,
we obtain $d^{*}{\simeq}1.8$ in the RH model with $W_{2}=1.0$.
The emergence of
this anomalous dimension $d^{*}$
may be due to the multifractal nature
of the zero-energy wavefunction
in the RH model \cite{lan}.

Y.H was supported in part by  Grant-in-Aid from the Ministry of Education,
Science and Culture of Japan.
The computation in this work has been done
using the facilities of the Supercomputer Center,
ISSP, University of Tokyo.

\begin{figure}

Fig. 1
Density of states
for the RS model (black) and the RH model (white),
where $W_{1}=2.3$, $W_{2}=1.0$ and $L^{2}=50^{2}$.
A finite width $\delta=0.02$ is given to the delta functions,
although
the results do not seriously depend on the small change of $\delta$.
The line is guide for eyes, which is $\sim |E|^{0.39}$.

Fig. 2
The Thouless number $g(E,L)$
for the RS model (black) and the RH model (white),
where $W_{1}=2.3$, $W_{2}=1.0$ and $L^{2}=50^{2}$.

Fig. 3
Scaling function $F$ and
the localization length ${\xi}(E)$ for the RS model.
The data are $g(E,L)$ near zero energy with $W_{1}=2.3$,
$L=30$, $40$ and $50$ and
different symbols correspond to different $L$'s.
The localization length $\xi (E)$ is
${\xi}(E)/{|E|^{\beta}}
=1+0.057{\epsilon}-0.78{\epsilon}^{2}
-0.32{\epsilon}^{3}+0.51{\epsilon}^{4}$,
where $\beta=0$
and $\epsilon =|E|$.
The scaling function is
${F}(y)=\log(y/230)(-0.12-0.00060y+0.0000032y^{2})$.
The curve beyond the data points is guide for eyes.

Fig. 4
Scaling function $F$ and
the localization length ${\xi}(E)$ for the RH model.
The data are $g(E,L)$ near the zero energy with $W_{2}=1.0$,
$L=30$, $40$ and $50$ and
different symbols correspond to different $L$'s.
${\xi}(E)/{|E|^{\beta}}
=1+0.47{\epsilon}-0.0057{\epsilon}^{2}
+0.20{\epsilon}^{3}-0.10{\epsilon}^{4}$,
where
$\beta=-0.75$
and $\epsilon =|E|$.
The scaling function is
${F}(y)=y^{-0.15}(0.44-0.0097y+0.000085y^{2})$.
The curve beyond the data points is guide for eyes.

\end{figure}


\begin{references}
\bibitem{2sc}
Here we fixed strength of randomness and
confirmed
$g(L,E,W)$'s with different $L$ and $E$
are on a universal curve $g=F(y)$.
The form of $g=F(y)$ and $\beta$
depend on the strength of randomness.
\end{references}
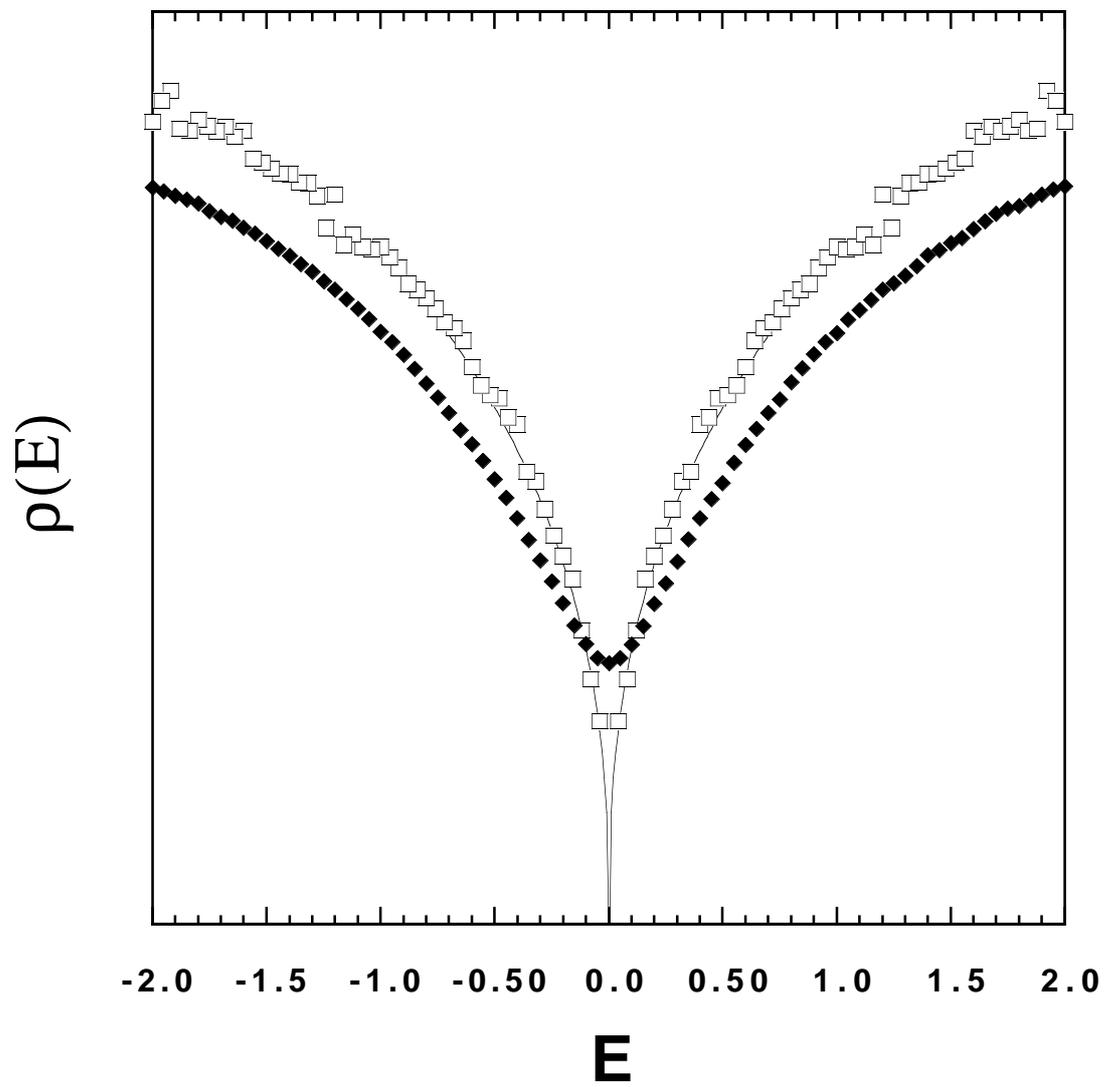
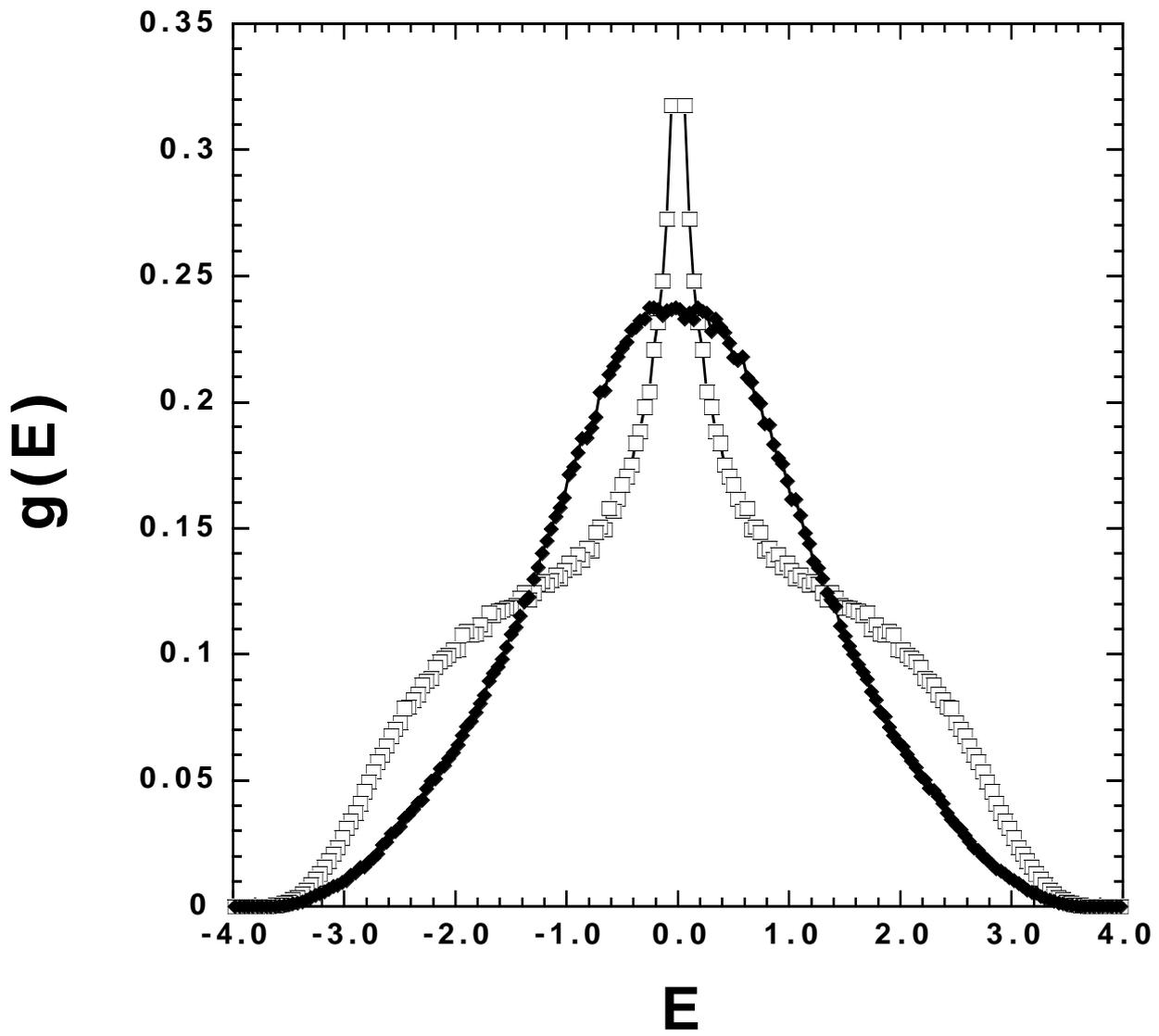
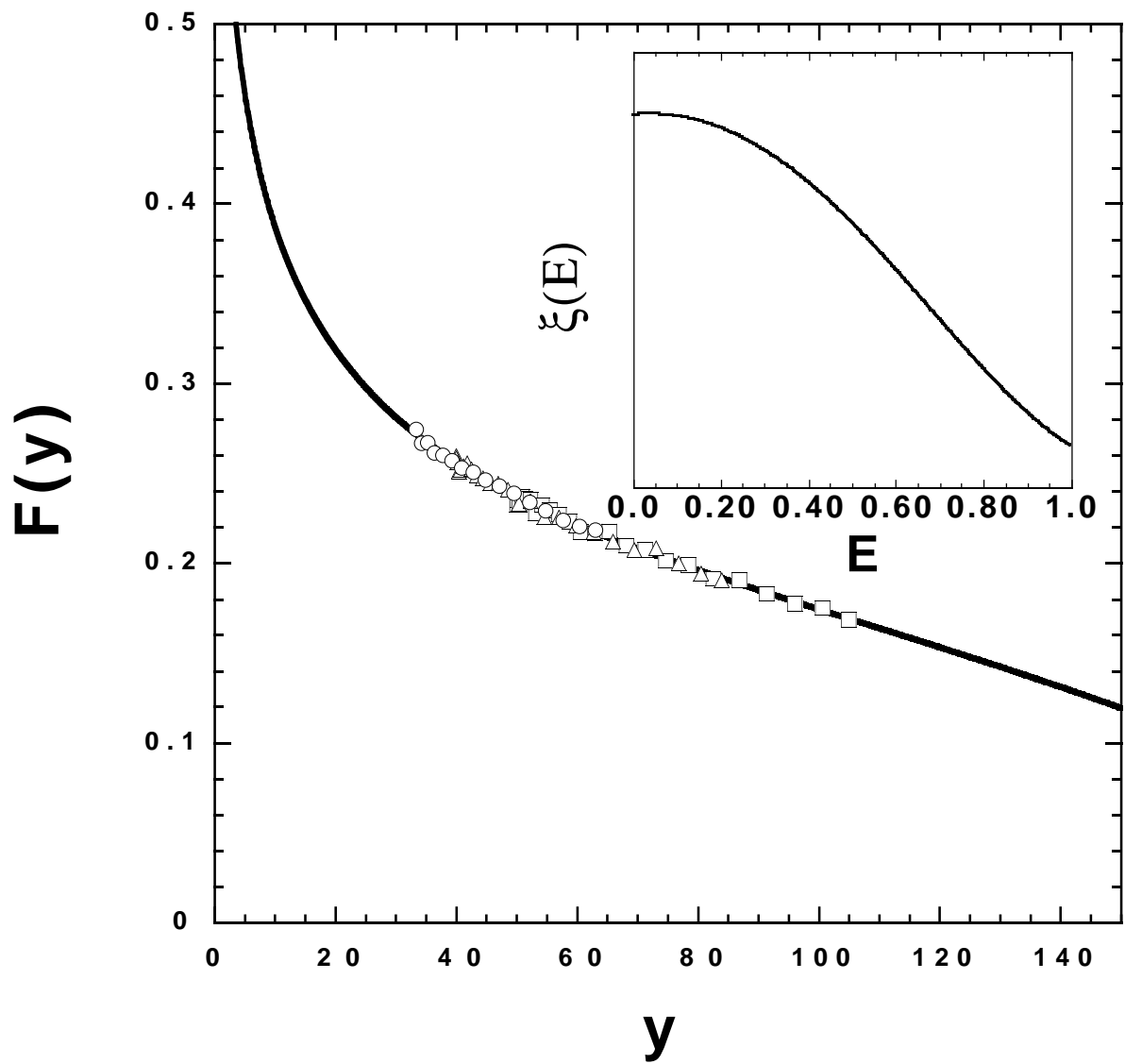
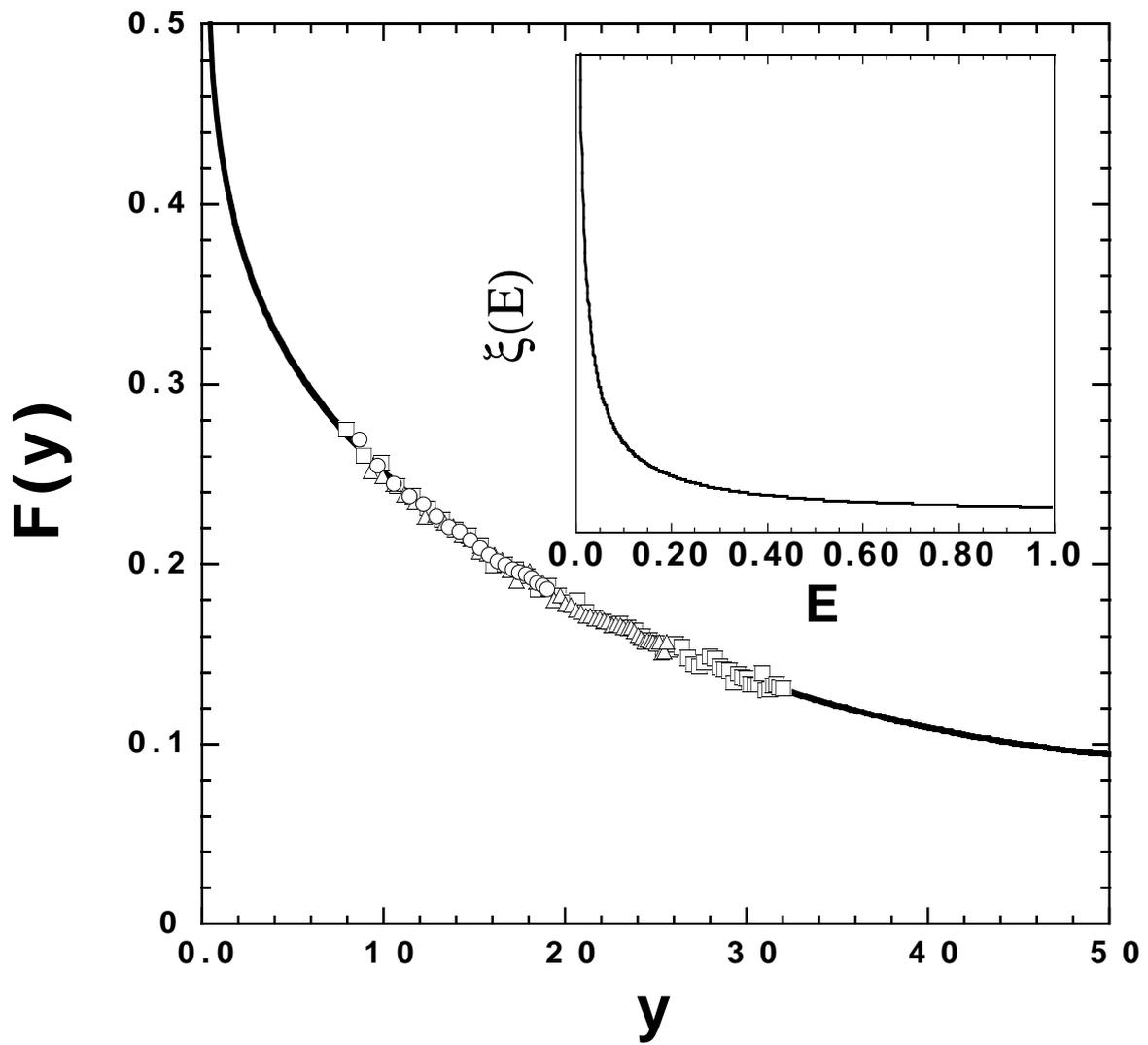
\end{document}